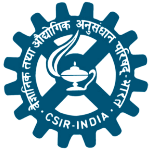

# Melting of Quarkonia in strong magnetic field

Manohar Lal, Siddhartha Solanki, Rishabh Sharma & Vineet Kumar Agotiya*

Department of Physics, Central University of Jharkhand Ranchi- 835 222, India

In this paper, spectra of the quarkonium states has been studied using the conditions temperature, chemical potential and the magnetic field. Here our main focus is to study the effect of strong magnetic field on the quarkonium properties. The binding energies and the dissociation temperature for the ground and the first excited states of the charmonium and bottomonium in the presence of strong magnetic field at chemical potential $\mu$ = 500 $MeV$ has been studied. Here we use quasiparticle(QP) Debye mass depending upon temperature, magnetic field and chemical potential obtained from the quasiparticle approach. The Debye mass strongly increases at different values of temperature and magnetic field. The binding energy decreases with increase in the temperature at different magnetic field eB=0.3, 0.5, and 0.7 $GeV^2$ and also decreases with magnetic field at different at T=200,300 and 400 $MeV$ for the $J/\psi$, $\Psi'$, $\Upsilon$, and $\Upsilon'$ states of the quarkonia. The dissociation temperature of the quarkonium states falls with the increasing values of the magnetic field at critical temperature $T_c$=197 $MeV$.

**Keywords**: Strongly Interacting Plasma; Dissociation Temperature; Quasi Particle Debye Mass; Magnetic Field; Quark-gluon Plasma; Chemical Potential; Heavy Ion Collision

## 1 Introduction

Ongoing experiment at Large Hadron Collider (LHC), Relativistic Heavy ion collision(RHIC) at CERN, Switzerland and Brookhaven National Laboratory(BNL), USA are capable of recreating the universe in the laboratory by making either nucleus-nucleus collision or by colliding the proton with the nucleus at various center of mass energy $\sqrt{S_{NN}}$. These collisions confirmed the fourth state of the matter known as the Quark Gluon Plasma(QGP). Under the extreme conditions of the temperature, chemical potential etc. the study of nuclear matter becomes subject of research interest in modern science for the last five past decades. Most of the studies have confirmed the non-central collision[1]. In the heavy-ion collision (HICs), a large magnetic field is produced when the heavy ions are in relative motion. The magnetic field so produced is in the direction perpendicular to the direction of the reaction plane. The magnitude of the produced magnetic field at an early stage of the universe is found to be very large. But this magnetic field decays immediately after the collision as it is inversely proportional to the square of the time[2-3]. To study the effect of the external anisotropic field, we need to modify the present theoretical models as well as the experimental techniques so, that the properties of the quark-gluon plasma can be easily studied. From the last decade, the major focus is to study the effect of magnetic field on the quark gluon properties. There are number of phenomenon arises because of the presence of the magnetic field. Some major are chiral magnetic effect[4], finite-temperature magnetic catalysis[5] and inverse magnetic catalysis[6], chiral and color symmetry broken/restoration phases[7], synchrotron radiation[8] and the dilepton production from the hot magnetized quantum chromodynamics (QCD) plasma[9] and also its effect can be seen in strongly coupled plasma[10]. PHENIX Collaboration provides experimental shreds of evidence for the external anisotropy which challenges the existing theoretical results[11]. After PHENIX results[11], several modifications have been performed to investigate the QGP properties in the presence of the strong magnetic field in non-central heavy-ion collisions(HICs)[12]. Since the heavy quarks pairs produced at ultra-relativistic heavy-ion collision (URHICs) has a very short time scale of $1/2m_Q$, $m_Q$ is the mass of the heavy quark whether it is charm quark or bottom quark which is very close to the scale of magnetic field generated during the time of heavy-ion collisions. Therefore, it becomes worthy to study the properties of the heavy quarkonia in the strong magnetic field regime. The formation time of the quark-antiquark pair is 1−2fm depending upon momenta and their resonances. To study the spectral properties of the quarkonia, potential model based on several parameters like temperature T, distance scale

---

*Corresponding author: (Email: agotiya81@gmail.com)

'r', chemical potential $\mu$, and magnetic field $eB$ etc. are of utmost important. Potential model studies are easy to carry out in comparison to the Lattice-based phenomenology[13]. Real part of the potential is screened due to Debye screening (Debye mass) and the imaginary part gives the thermal width of quark-antiquark resonance[14]. The heavy quarkonia meson spectroscopy in the strong magnetic field has been investigated using 3$D$ Harmonic oscillator and Cornell potential with spin effect in[15]. Also in[16], the effect of constant magnetic field on the quarkonia potential, finite temperature and the Debye screening has been investigated. Recently, the effect of the strong magnetic field on the quark-antiquark potential in finite temperature has been explored by modifying the real part of the potential[17] in the thermal QCD and the properties of the QGP has been studied. Dissociation of the heavy quarkonia due to the Landau damping in the strong magnetic limit using the complex heavy quark potential has been studied in[18-19]. In the presence of the magnetic field there is anisotropy in the medium and the anisotropic interactions has been found in[20]. Because of the oppositely directed motion of the heavy quark ions, a strongly transient magnetic field produced during the off central collision as suggested by[21]. Studies[22-23] provides unique environment to study the quarkonia in the presence of magnetic field. For the better understanding of the quark gluon plasma, the background magnetic field plays a significant role. Particularly this strong magnetic field in the Hot QCD changes the thermodynamically properties of the quark-gluon plasma such as the energy density, pressure, speed of the sound $C_s^2$ etc.[24]. Due to this magnetic field several other macroscopic properties such as transport coefficient e.g. viscosity, electrical conductivity will also be modified in[25]. In the present paper our prime focus is to study the quarkonium properties in the presence of the strong magnetic field $eB >> T^2$ at constant value of the chemical potential. We use the chemical potential $\mu = 500 MeV$ throughout the manuscript. By using the effective quasiparticle model purposed by Ravi Shankar and Chandra[26-27], we modify the magnetized Debye mass into another form of the Debye mass which depends upon the temperature, chemical potential and the magnetic field. It should be noted that the effect of the chemical potential has been introduced through the two-loop coupling constant. Then using this potential model, the binding energy, dissociation temperature has been calculated.

## 2 Heavy Quarkonia potential

The potential-based phenomenology is one of the most important mathematical model to study the spectra of the quarkonia states. Heavy quarkonia potential is modified[28] through the Fourier transform and then taking the inverse Fourier transform, we get the potential in the limiting case r $>> \frac{1}{m_D}$,

$$V(r, T, \mu, eB) \approx -\frac{2\sigma}{m_D^2(T,\mu,eB)r} - \alpha m_D(T, \mu, eB) \quad \ldots (1)$$

where σ is the string tension and its value is taken as 0.184GeV$^2$ whereas $\alpha$ is the two loop coupling constant depending upon temperature and chemical potential, eB is the magnetic field, μ is the chemical potential, T is the temperature, $m_D^2$ is the quasiparticle Debye mass and 'r' is the distance scale.

## 3 Debye Mass from the Quasi-Particle Model in the Presence of Magnetic Field

Since it is well known that the plasma contains both the charged and the neutral quasi-particle, hence it shows collective behavior. The Debye screening length is an important quantity used to measure the effect of electric potential on the quark-gluon plasma. The screening effect of color forces in QGP can be fully described in terms of Debye mass. One can find the conventional definition of the Debye mass from[29]. The Debye mass in the presence of strong magnetic field for $N_f = 3$ has been found in[30]. The Debye mass in the presence of chemical potential and the magnetic field can be written as

$$m_D^2(T, \mu, eB) = 4\pi\alpha_s \left[T^2 + \frac{3eB}{2\pi^2}\right] \quad \ldots (2)$$

The QCD coupling constants $\alpha_s (\mu, T)$ at finite chemical potential and temperature[31-32] is given as

$$\alpha_s(\mu, T) = \frac{6\pi}{(33-2N_f)\$} \left(1 - \frac{3(153-19N_f)}{(33-2N_f)^2} \frac{\ln(2\$)}{\$}\right)$$

where $\$ = \ln\left(\frac{T}{\lambda_T}\sqrt{1+\frac{\mu^2}{\pi^2 T^2}}\right)$

where σ is the string tension and here we take σ= 0.184GeV$^2$ whereas $\alpha$ is the two loop coupling constant depending upon temperature and chemical potential, μ is the chemical potential, T is temperature which is taken as 200, 300 and 400MeV, $\lambda_T$ is the QCD coupling scale and $N_f = 3$ is the number of flavor.

Debye mass parametrizes the dynamically generated screening of chromo-electric field in Hot

QCD. The Debye mass for QED plasma and non-perturbative gauge-invariant vector-like theories at zero chemical potential can be found in[33-34].

## 4 The Binding Energy and the Dissociation Temperature

Since the potential defined by Eq. (1) mapped as $1/r$ which is quite similar to the hydrogen atom problem. So, we solve the Schrodinger equation to study spectrum of the quarkonium states in the hot QGP medium. Since the binding energy of the bound states of the quarkonia at $T = 0$ can be defined as the difference of energy between the mass of quarkonia $m_Q$ and the bottom or charm threshold. From the literature[35], the binding energy at finite temperature is the distance between the continuum threshold and the peak position The solution of the Schrodinger equation for the potential defined by the Eq. (1) gives the energy eigenvalues for the ground and the excited states of the charmonium and bottomonium $J/\psi$, $\psi'$, $\Upsilon$ and $\Upsilon'$ as:

$$E_n = -\frac{m_Q \sigma^2}{n^2 m_D^4(T_D)} \quad \ldots (3)$$

where $m_Q$ is the mass of the quarkonium states i.e the charomonia and bottomonium, n is number of the energy levels and σ is string tension which is taken as 0.184GeV$^2$.

The binding energy of charmonium and bottomonium state at particular values of temperature becomes smaller or equal to the value of mean thermal energy; i.e. the state of quarkonia is said to be dissociated at that given value of temperature. The binding energy of the bound states of quarkonium i.e. equal to the mean value of the thermal energy as given below:

$$E_n = \frac{m_Q \sigma^2}{n^2 m_D^4(T_D)} = \begin{cases} T_D \text{ for upper bound state} \\ 3T_D \text{ for lower bound state} \end{cases} \quad \ldots (4)$$

where $m_Q$ is the mass of the quarkonium states i.e the charmonium and bottomonium, n is number of the energy levels and σ is string tension which is taken as 0.184GeV$^2$.

Literature survey reveals that there are several methods adopted for calculating the dissociation temperature as in[36-38].

Dissociation temperature has also been calculated using the thermal width[35]. It has also been found in[28], the thermal width is equal to twice the real part of the binding energy. The dissociation temperature of the quarkonium states can also be obtained by using the condition $E_B = 3T_D$ and $E_B = T_D$ for the lower and the upper bound respectively. Here we calculate the dissociation temperature for the lower bound and the upper bound of the quarkonium states at various values of the magnetic field and fixed value of the temperature $T_c$ = 197 MeV. The dissociation temperature for the lower bound ($E_B = 3T_D$) and the upper bound ($E_B = T_D$) of charmonium and the bottomonium states at different magnetic field has been given in Table- 1 & 2 respectively.

## 5 Results and Discussion

In the present work, we have studied the properties of the quarkonia in the presence of strong magnetic field at a constant value of chemical potential. Here we use the Debye mass depending upon the temperature, chemical potential, and magnetic field obtained from the quasiparticle model. It should be noted that we employed the two-loop coupling constant which depends upon the temperature and the chemical potential. For studying the behavior of the magnetic field on the quarkonium states, we use the T=200, 300, and 400 MeV and eB= 0.3, 0.5, and 0.7 GeV$^2$. However, these values of the temperature and magnetic field are taken arbitrarily for studying the spectra of the heavy quarkonia states. Fig. 1 shows the variation of potential with 'r'(fm) at different values of temperature and the variation of potential with 'r'(fm) at different values of magnetic field has been shown in Fig. 2. It has been observed that with the increase in the temperature and magnetic field, there is an increase in the potential with the 'r'(fm). Whereas Fig 4 shows the variation of the Debye mass

Table 1 — The dissociation temperature (is in unit of $T_c$) for the lower bound state (B.E =3$T_D$) with $T_c$ = 197*MeV* for the different states of quarkonia at fixed value of $\mu$ = 500*MeV*.

| State | eB=0.3GeV$^2$ | eB=0.5GeV$^2$ | eB=0.7GeV$^2$ |
|---|---|---|---|
| $J/\psi$ | 1.916243 | 1.662436 | 1.408629 |
| $\Upsilon$ | 2.296954 | 2.124394 | 1.916343 |
| $\psi'$ | 1.269876 | 1.154822 | - |
| $\Upsilon'$ | 1.789340 | 1.535532 | 1.281725 |

Table 2 — The dissociation temperature (is in unit of $T_c$) for upper bound state (B.E =$T_D$) with $T_c$ = 197*MeV* for the different states of quarkonia at fixed value of $\mu$ = 500*MeV*.

| State | eB=0.3GeV$^2$ | eB=0.5GeV$^2$ | eB=0.7GeV$^2$ |
|---|---|---|---|
| $J/\psi$ | 2.296954 | 2.043147 | 1.789340 |
| $\Upsilon$ | 2.677664 | 2.550761 | 2.296954 |
| $\psi'$ | 1.789340 | 1.535532 | 1.281725 |
| $\Upsilon'$ | 2.170050 | 1.916243 | 1.662436 |

with the magnetic field at various temperatures. The Fig. 3 shows the behavior of the Debye mass with the temperature at the various magnetic field. It has been observed that the Debye mass increases with increase in the temperature as can be seen from Fig. 4 and also with the magnetic field as we observed from Fig. 3. Figs. 5,6,7,8,9,10,11 and 12 shows the variation of the binding energy of the quarkonium states such as $J/\psi$, $\Psi'$, $\Upsilon$, and $\Upsilon'$. The Fig. 6, Fig. 8, Fig. 10, and Fig. 12 shows the behavior of the binding energy of quarkonia states $J/\psi$, $\Upsilon$, $\Psi'$, and $\Upsilon'$ with magnetic field at different temperatures T = 200,300 and 400MeV respectively. On the other hand, the behavior of the binding energies of the charmonium and bottomonium states with tilde $T/T_C$ at different values of magnetic fields eB = 0.3, 0.5, and 0.7 $GeV^2$ has been shown in the Fig. 5, Fig. 7, Fig. 9 and Fig. 11. It has been observed from the figures that the binding energies for the states of the charmonium and bottomonium decrease with the temperature as well as with the magnetic field at constant chemical potential. It should also be noted from the Figs. 5,6,7,8,9,10,11 and 12 that in comparison to the binding energy of quarkonium states at various temperature, there is a strong decrease in the binding energy of these states in the presence of the strong magnetic field. This type of behavior for the binding energy has also been seen in[39] at a fixed value of magnetic field eB = $15 m_\pi^2$ at

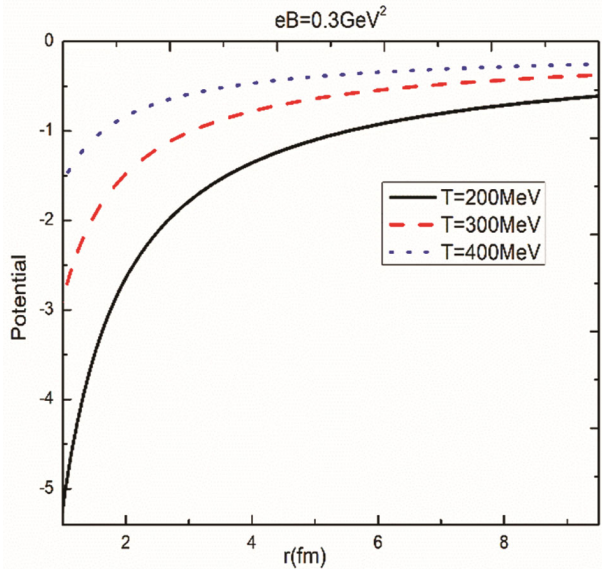

Fig. 1 — The variation of potential with r (fm) at different values of temperature at constant value of magnetic field eB=0.3 GeV$^2$.

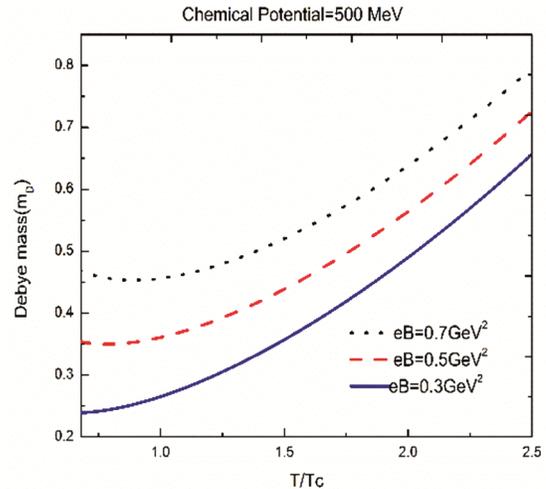

Fig. 3 — The variation of quasi-particle Debye mass with temperature at different values of magnetic field.

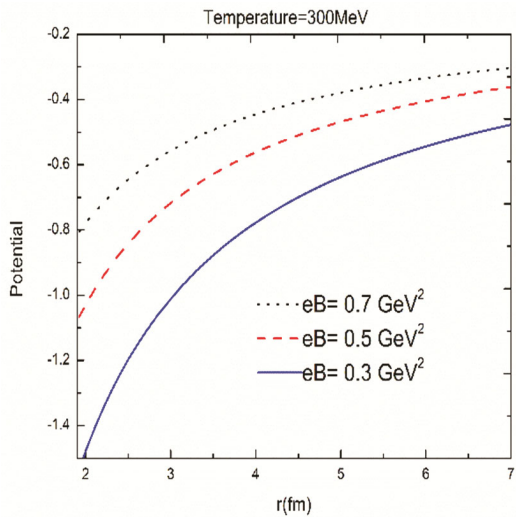

Fig. 2 — The variation of potential with r (fm) at different values of magnetic field at constant value of temperature T=300 MeV.

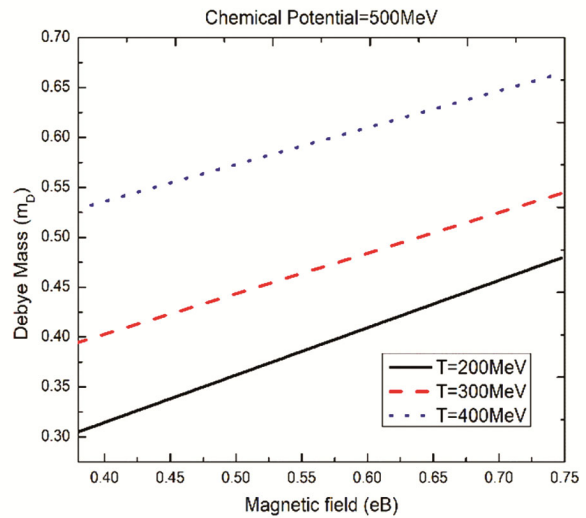

Fig. 4 — The variation of quasi-particle Debye mass with the magnetic field at different values of temperature.

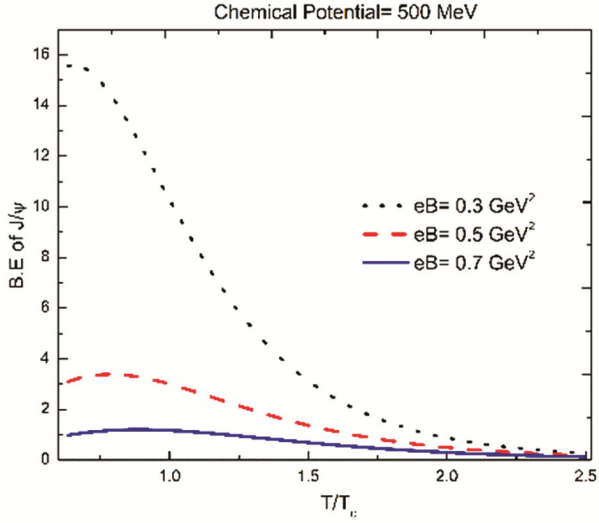

Fig. 5 — The variation of Binding energy (B.E) of *J/ψ* with temperature at different values of magnetic field.

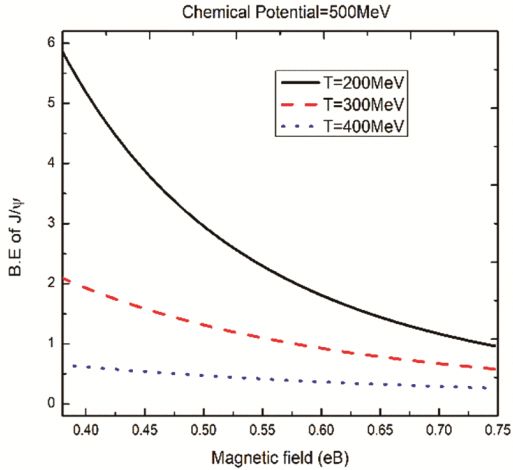

Fig. 6 — The variation of Binding energy (B.E) of *J/ψ* with the magnetic field at different values of temperature.

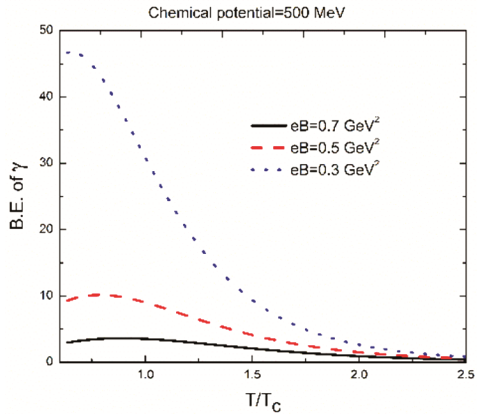

Fig. 7 — The variation of Binding energy (B.E) of upsilon (ϒ) with temperature at different values of magnetic field.

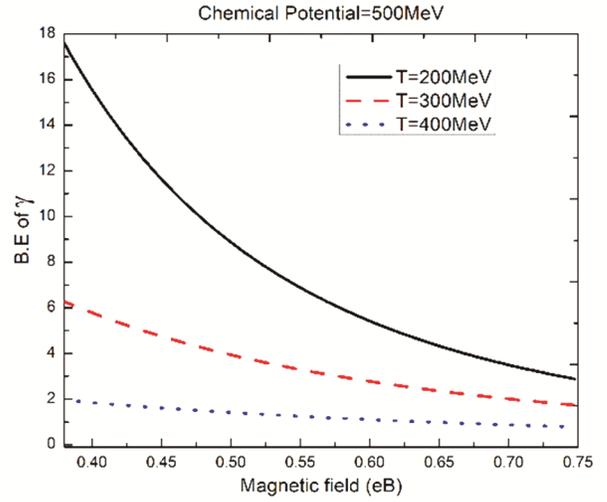

Fig. 8 — The variation of Binding energy (B.E) of upsilon (ϒ) with the magnetic field at different values of temperature.

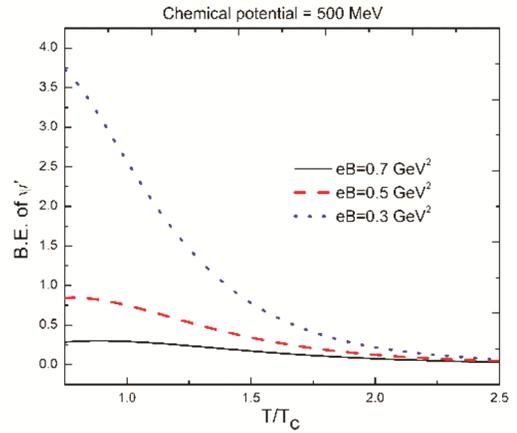

Fig. 9 — The variation of Binding energy (B.E) of *Ψ'* with temperature at different values of magnetic field.

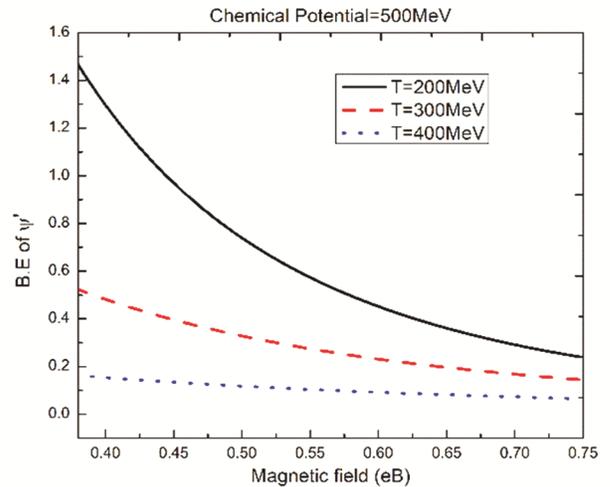

Fig. 10 — The variation of Binding energy (B.E) of *Ψ'* with the magnetic field at different values of temperature.

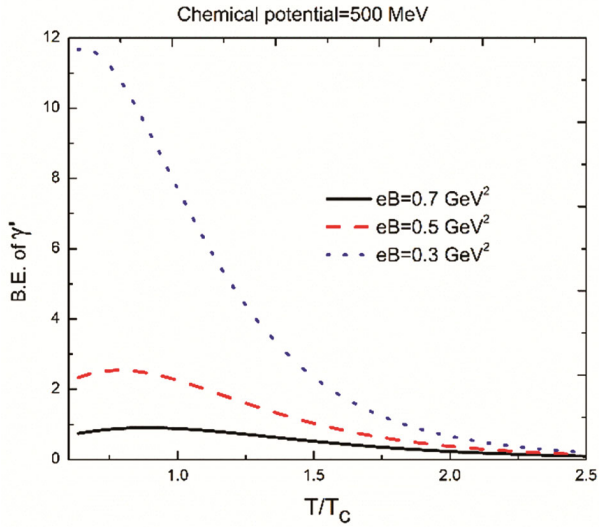

Fig. 11 — The variation of Binding energy (B.E) of upsilon prime (Υ'') with temperature at different values of magnetic field.

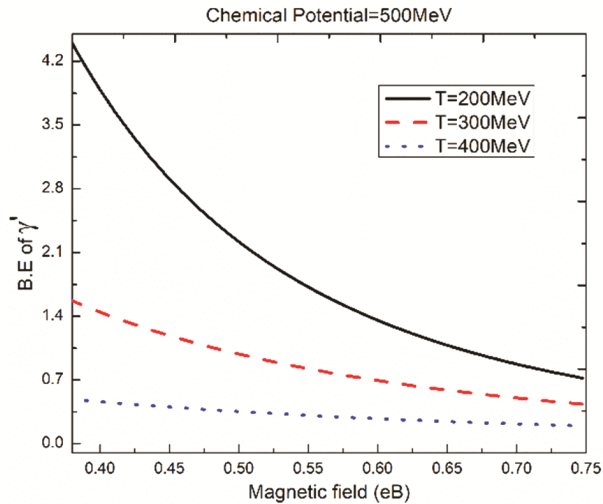

Fig. 12 — The variation of Binding energy (B.E) of upsilon prime (Υ'') with the magnetic field at different values of temperature.

varying chemical potential. We have also calculated the dissociation temperature for the charmonium and bottomonium states in the presence of strong magnetic field. Here the value of $T_C$ = 197 MeV is kept fixed while calculating the dissociation temperature. Table-1 shows the $T_D$ for the lower bound and Table-2 shows the $T_D$ for the upper bound of various quarkonium states at various values of the magnetic field 0.3, 0.5, and 0.7 $GeV^2$. For the lower bound and the upper bound, the dissociation temperature for the quarkonium states decreases with the increasing magnetic field at constant chemical potential.

The dissociation temperature for the $J/\psi$ is found to be 2.29, 2.04, and 1.78 in terms of $T_c$ at eB = 0.3, 0.5, and 0.7 $GeV^2$. Similarly, for Υ it is found to be 2.6, 2.5, and 2.2 in terms of $T_c$ at eB = 0.3, 0.5, and 0.7 $GeV^2$ for the upper bound state at $T_c$ =197MeV. The decreasing pattern for the dissociation temperature of the other states of the quarkonium can be seen in Table- 1 & 2. It is interesting to know that as we go from $J/\psi$ to Υ and $\Psi$' to Υ'' at every increasing value of the magnetic field, the dissociation temperature increases. This implies that the magnetic field plays a significant to understanding the dynamic properties of the quark-gluon plasma at constant chemical potential.

## 6 Conclusion and the Future outlook

The decrement in the binding energy accounts for the stronger nature of quark-antiquark potential in the presence of a strong magnetic field. This means that the higher excited state of the quarkonia melts at the lower dissociation temperature with the increasing magnetic field, so the ground states of quarkonia are of utmost important. Properties of the quarkonia can be easily studied in the presence of the magnetic field. This is because the bound states of the quarkonia take more time to dissociate and hence it is feasible to study all the interaction effects in the higher-order states of the quarkonium. Hence it can be concluded that quarkonium states take more time to melt, this correspond greater probability to study the dynamical properties of the quarkonium states. There are some other useful parameters like electrical conductivity which increases with the magnetic field and hence the formation time of the fireball (QGP) increases and dynamical properties of the quark gluon plasma (QGP) could be easily understood. It is also noteworthy to mention that with the increasing value of the magnetic field up to the scale $1 GeV^2$ as can be seen from the literature, similar behavior has been observed for the binding energy and the dissociation temperature of the quarkonium states at chemical potential $\mu$=500MeV. The dissociation temperature obtained in the presence of a magnetic field could be employed for calculating the equation of states EoS and the suppression of the quarkonium states.

## Acknowledgement

VKA acknowledges the Science and Engineering Research Board (SERB) Project No. EQ/2018/000181 New Delhi for providing financial support. We record our sincere gratitude to the people of India for their generous support for the research in basic sciences.